\documentstyle[aaspp4]{article}

\received{}
\accepted{}
\journalid{VOL}{JOURNAL DATE}
\articleid{START PAGE}{END PAGE}
\paperid{MANUSCRIPT ID}

\lefthead{Yoshida}
\righthead{$r$-modes of non-isentropic relativistic stars}

\begin{document}


\title{$r$-modes of slowly rotating non-isentropic relativistic stars}
\author{Shijun Yoshida\altaffilmark{1}}
\affil{Astronomical Institute, Graduate School of Science, 
Tohoku University, Sendai 980-8578, 
Japan \\ yoshida@astr.tohoku.ac.jp}

\altaffiltext{1}{Research Fellow of the Japan Society for 
the Promotion of Science.}

\begin{abstract}

We investigate properties of $r$-modes characterized by regular 
eigenvalue problem in slowly rotating relativistic polytropes. 
Our numerical results suggest that discrete $r$-mode solutions for the 
regular eigenvalue problem exist only for restricted polytropic models. 
In particular the $r$-mode associated with $l=m=2$, which is considered 
to be the most important for gravitational radiation driven instability, 
do not have a discrete mode as solutions of the regular eigenvalue 
problem for polytropes having the polytropic index $N > 1.18$ even in 
the post-Newtonian order. 
Furthermore for a $N=1$ polytrope, which is employed as 
a typical neutron star model, discrete $r$-mode solutions for regular 
eigenvalue problem do not exist for stars whose relativistic factor 
$M/R$ is larger than about $0.1$. Here $M$ and $R$ are  
stellar mass and stellar radius, respectively.    
 
\end{abstract}



\section{Introduction}

Since the discovery of the gravitational radiation driven instability of
the $r$-modes by Andersson (1998) and Friedman \& Morsink (1998),
a large number of studies on the properties of $r$-modes and inertial modes 
of rotating stars have been done to prove their possible 
importance in astrophysics (for a review see, e.g., Friedman \& 
Lockitch 1999; Andersson \& Kokkotas 2000). Although early investigations  
were mainly applied to young hot neutron stars (e.g., Lindblom, Owen, \& 
Morsink 1998; Owen et al. 1998), recently some authors have also discussed 
possible roles of the $r$-mode instability in old and cool neutron stars 
with a solid crust and a magnetic field, such as those found in low mass 
X--ray binaries (LMXBs) (e.g., Andersson, Kokkotas, \& Stergioulas 1999;
Bildsten \& Ushomirsky 2000; Rezzolla, Lamb, \& Shapiro 2000; Yoshida \& Lee 
2001).

Although our understandings of $r$-modes have been largely improved by recent 
energetic investigations as mentioned above, most studies concerning $r$-modes 
have been done in the framework of Newtonian gravity. 
As for $r$-modes in the framework of general relativity,  
Kojima (1998) found possible existence of continuous spectrum for pure 
$r$-mode in relativistic stars. 
Here throughout this paper we call the rotational mode whose eigenfunctions 
are composed of only one axial parity component in the non-rotating star limit 
``pure $r$-mode''.
Beyer \& Kokkotas (1999) generally 
verified the existence of a continuous spectrum. 
Kojima's formalism was developed to that with high order 
rotational effects by Kojima \& Hosonuma (1999; 2000). 
Very recently, Lockitch, 
Andersson, \& Friedman (2001) have shown that pure $r$-modes in a 
relativistic star can exist only in non-isentropic stars, whose 
specific entropy distribution of a star is not constant. 
As for rotation induced modes in isentropic stars, they have shown that modes 
similar to inertial modes, namely hybrid modes, in a Newtonian rotating 
star (see, e.g., Lockitch \& Friedman 1999; Yoshida \& Lee 2000), whose 
eigenfunctions are composed of both polar and axial parity components, 
exist in uniform density stars (see also Lockitch 1999).

In the study by Lockitch et al. (2001), they derived the basic equation 
for the pure $r$-mode in relativistic non-isentropic stars. Their equation 
is the same as that obtained by Kojima (1998). As mentioned by Kojima
(1998), this equation has a continuous part of frequency as its solution
when the system of equations is that of
the singular eigenvalue problem. Kojima 
(1998) also pointed out that the system of the equation becomes that of
the regular eigenvalue problem when the frequency domain is in some range. 
For such a regular eigenvalue problem, discrete 
eigenfrequencies may exist for the same equation as well as continuous 
ones. 
Lockitch et al. (2001) have studied the relativistic $r$-mode 
characterized by the regular eigenvalue problem. In their study discrete 
$r$-mode solutions for the regular eigenvalue problem in uniform density 
stars were obtained. It is reasonable to consider that such discrete 
$r$-mode solutions correspond to those of Newtonian $r$-mode.  
However their study was restricted to the case of uniform density 
stars. As mentioned in the next section, the frequency range for the 
regular eigenvalue 
problem is confined to the narrow region and depends on the structure of 
equilibrium stars. Furthermore, needless to say, relativistic effects 
on stellar oscillations in a neutron star are not negligible. 
Thus, it is interesting and important to answer to the 
question of whether $r$-mode solutions for the regular eigenvalue problem 
exist in all non-isentropic relativistic stars. 
In this paper, therefore, we study the $r$-mode characterized by 
regular eigenvalue problem for large class of polytropic and non-isentropic 
relativistic stellar models. 
The plan of this paper is as follows. We start, in \S 2, with brief 
description of our method of solution.  In \S 3, we describe 
the properties of the $r$-mode characterized by the regular eigenvalue 
problem in both the relativistic and the post-Newtonian stellar models.  
\S 4 is devoted for discussions and conclusions. Throughout this paper we 
will use units in which $c=G=1$, where $c$ and $G$ denote the velocity of 
light and the gravitational constant, respectively.


\section{Method of solutions}

We consider slowly rotating relativistic stars with a uniform angular 
velocity $\Omega$, where we take account of the first order rotational 
effect in $\Omega$. The geometry around the equilibrium stars can be 
described by the following line element (see, e.g. Thorne 1971): 
\begin{eqnarray}
ds^2 = - e^{2 \nu(r)} dt^2 + e^{2 \lambda(r)} dr^2 + r^2 d\theta^2 + 
           r^2 \sin^2 \theta d\varphi^2 - 
           2 \omega(r) \, r^2 \sin^2 \theta dt d\varphi  \, . 
\label{metric}
\end{eqnarray}
Once an equation of state, $p=p(\epsilon,s)$ for a perfect fluid is given, 
we can easily obtain the structure of the slowly rotating stars, where $p$, 
$\epsilon$, and $s$ denote pressure, mass-energy density and specific 
entropy, respectively. Throughout this paper, the polytropic equation of 
state is assumed:   
\begin{equation}
p = K\, \epsilon^{1+\frac{1}{N}} \, ,
\end{equation}
where $N$ and $K$ are constants.

We consider oscillation modes in rotating relativistic stars such that 
the eigenfunctions are stationary and are composed of only one axial parity 
component in the limit $\Omega\rightarrow0$. Subclass of those modes should 
be a relativistic counterpart of $r$-modes, which are able to oscillate in 
all slowly rotating Newtonian fluid stars. According to Lockitch et al. 
(2001), such modes are allowed to exist only if the star has non-isentropic 
structures. Therefore, we assume stars to be non-isentropic, although the 
effects due to deviation from isentropic structure on oscillation modes 
do not appear in the first order in $\Omega$. According to the formalism 
by Lockitch et al. (2001) (see, also, Kojima 1998), let us write down the 
pulsation equation for 
relativistic $r$-modes with accuracy up to the first order in $\Omega$. 
The metric perturbation $\delta g_{\alpha\beta}$ and the Eulerian changes  
of the fluid velocity $\delta u^\alpha$ that do not vanish in the limit 
$\Omega\rightarrow0$ are given as 
\begin{equation}
(\delta g_{t\theta},\delta g_{t\varphi})=i h_{0,l}(r) \, 
\left(-\frac{\partial_\varphi Y_{lm}(\theta,\varphi)}{\sin\theta}\, ,
\sin\theta\partial_\theta Y_{lm}(\theta,\varphi)\right)\, e^{i \sigma t} \, , 
\label{ex-g}
\end{equation}
\begin{equation}
(\delta u^\theta,\delta u^\varphi)=\frac{i U_{l}(r)}{r^2} \, 
\left(-\frac{\partial_\varphi Y_{lm}(\theta,\varphi)}{\sin\theta}\, ,
\frac{\partial_\theta Y_{lm}(\theta,\varphi)}{\sin\theta}\right)\, 
e^{i \sigma t} \, ,
\label{ex-v}
\end{equation}
where $Y_{lm}(\theta,\varphi)$ are usual spherical harmonic functions, and 
$\sigma$ denotes oscillation frequency measured in the inertial frame. 
All other perturbed quantities become higher order in  
$\Omega$. 
Note that 
this form of eigenfunctions is the same as that for zero-frequency modes 
in a spherical non-isentropic star, because $r$-modes become zero-frequency 
ones in the limit $\Omega\rightarrow0$.(Thorne \& Campolattaro 1967)  
The metric perturbation $h_{0,l}$ obeys a second order ordinary 
differential equation,   
\begin{eqnarray}
&&D_{lm}(r;\bar{\sigma}) \, 
\left[e^{\nu-\lambda}\frac{d}{dr}
\left(e^{-\nu-\lambda}\frac{dh_{0,l}}{dr}\right)-\left(\frac{l(l+1)}{r^2}+
\frac{ -2 + 2 e^{-2 \lambda}}{r^2}+8 \pi (p+\epsilon)\right)h_{0,l}\right] 
\nonumber \\ 
 &+& 16 \pi (p+\epsilon) h_{0,l}=0 \, ,
\label{b-eq}
\end{eqnarray}
where 
\begin{equation}
D_{lm}(r;\bar{\sigma})\equiv 1-\frac{2 m \bar{\omega}}{l(l+1)\bar{\sigma}} \,
,
\label{coe1}
\end{equation}
Here, we have introduced the effective rotation angular velocity of fluid,  
\begin{equation}
\bar{\omega} = \Omega - \omega \, , 
\end{equation}
and the corotating oscillation frequency, 
\begin{equation}
\bar{\sigma} = \sigma + m \Omega \, .
\end{equation}
The velocity perturbation of fluids $U_l$ is determined from the function 
$h_{0,l}$ through the following algebraic relation, 
\begin{equation}
\left[1-\frac{2 m \bar{\omega}}{l(l+1)\bar{\sigma}}\right]\, U_{l}+h_{0,l}=0 
\, . 
\label{b-eq2}
\end{equation}
Equations (\ref{b-eq}) and (\ref{b-eq2}) are our basic equations, which 
were derived by Kojima (1998) in his early work. 
Note that equations 
(\ref{b-eq}) and (\ref{b-eq2}) lose the meaning for the $l=0$ case, because 
there are no axial modes with $l=0$.  

Because equations (\ref{b-eq}) are second order ordinary differential 
equations, 
two boundary conditions are required to determine solutions uniquely. 
For physically acceptable solutions, the function 
$h_{0,l}$ must vanish both at the center and at the spatial infinity. 
Thus, boundary conditions are given as
\begin{eqnarray}
r \, \frac{d h_{0,l}}{dr} &-&(l+1)\, h_{0,l} = 0 \ \ \ \ 
{\rm as} \ r\rightarrow 0 \, , 
\label{bc1} \\
r \, \frac{d h_{0,l}}{dr} &+& l\, h_{0,l} = 0 \ \ \ \ 
{\rm as} \ r\rightarrow\infty \, . 
\label{bc2}
\end{eqnarray}
When those boundary conditions are imposed, our basic equations are solved 
as an eigenvalue problem with an eigenvalue $\bar{\sigma}$. As shown by 
Kojima (1998), when $D_{lm}(r;\bar{\sigma}) = 0$ is satisfied inside the 
star the system of our basic equations becomes a singular eigenvalue 
problem, because the function $D_{lm}(r;\bar{\sigma})$ is proportional to 
the coefficient of the second derivative of function $h_{0,l}$. 
On the other hand, we can treat the equation as a regular 
eigenvalue problem if $D_{lm}(r;\bar{\sigma}) = 0$ is not satisfied inside 
the star, because the last term of equation (\ref{b-eq}) vanishes outside the 
star as well as the function $D_{lm}(r;\bar{\sigma})$. 
Due to this singular property of the basic equation, Kojima also 
concluded continuous spectrum may exist in relativistic rotating stars. 
Beyer \& Kokkotas (1999) gave a mathematically rigorous proof for the 
existence of a continuous part within the spectrum. Recently Lockitch et al. 
(2001) have showed that the equality $D_{lm}(r;\bar{\sigma}) = 0$ must be 
satisfied at some spatial point for the equations to have non-trivial 
solutions $h_{0,l}$. 
Because $\bar{\omega}$ is a monotonically increasing positive function of $r$, 
according to the considerations of Lockitch et al. (2001), 
in order for the equation (\ref{b-eq}) to be treated 
as a regular eigenvalue problem
an frequency $\bar{\sigma}$ must be in a region,     
\begin{equation}
\frac{2 m \bar{\omega}(R)}{l(l+1)} < \bar{\sigma}  \le
\frac{2 m \bar{\omega}(\infty)}{l(l+1)} = 
\frac{2 m \Omega}{l(l+1)} \, , 
\label{dis-sp}
\end{equation}
where $R$ is stellar radius in the coordinate (\ref{metric}). Note that 
$\frac{2 m \Omega}{l(l+1)}$ is well-known Newtonian $r$-mode frequency. 
On the other hand, when $\bar{\sigma}$ satisfies an inequality,   
\begin{equation}
\frac{2 m \bar{\omega}(0)}{l(l+1)} \le \bar{\sigma} \le 
\frac{2 m \bar{\omega}(R)}{l(l+1)} \, , 
\label{con-sp}
\end{equation}
we must treat equation (\ref{b-eq}) as singular eigenvalue problem. 
Note that since $\bar{\omega}$ becomes $\Omega$ in Newtonian limit, we 
can easily see from inequalities (\ref{dis-sp}) and (\ref{con-sp}) that 
frequency in a range for existing non-trivial solution of equation 
(\ref{b-eq}) is getting closer to Newtonian $r$-mode frequency, 
$\frac{2 m \Omega}{l(l+1)}$, as $M/R\rightarrow0$.    
Lockitch et al. (2001) also found numerically discrete $r$-mode solutions 
with frequencies in range (\ref{dis-sp}) in uniform density relativistic 
stars. In this paper we will extend the study on the discrete $r$-mode 
solutions by Lockitch et al. (2001) to those for polytropic equation of 
states.

Next let us consider a post-Newtonian approximation of relativistic $r$-mode 
oscillation, by using a similar method to that of Lockitch et al. (2001). 
In this approximation, physical quantities of an equilibrium state can be 
written as 
\begin{equation}
\epsilon = \frac{\epsilon_0 \, \Theta^N(r)}{R^2}\,\frac{2 M}{R} + 
O\left(\left(\frac{2 M}{R}\right)^2\right) \, , 
\label{pn1}
\end{equation} 
\begin{equation}
p = \frac{p_0 \, \Theta^{N+1}(r)}{R^2}\,\left(\frac{2 M}{R}\right)^2 +
O\left(\left(\frac{2 M}{R}\right)^3\right) \, , 
\label{pn2}
\end{equation} 
\begin{equation}
e^{\nu} = 1 + O\left(\frac{2 M}{R}\right) \, , 
\label{pn3}
\end{equation} 
\begin{equation}
e^{\lambda} = 1 + O\left(\frac{2 M}{R}\right) \, ,
\label{pn4}
\end{equation} 
\begin{equation}
\bar{\omega} = \Omega \left(1 - \omega_0(r) \,\frac{2 M}{R} \right) + 
O\left(\left(\frac{2 M}{R}\right)^2\right) \, , 
\label{pn5}
\end{equation} 
where $\Theta$ is a Newtonian Lane-Emden function (see, e.g., Kippenhahn \& 
Weigert 1990) and $M$ denotes total mass of stars. 
Here $\epsilon_0$ and $p_0$ are constants determined from a 
solution of Lane-Emden equation. Function $\omega_0$ is given as solution 
of a differential equation, 
\begin{equation}
\frac{d^2\omega_0(r)}{dr^2}+\frac{4}{r}\,\frac{d\omega_0(r)}{dr}+
16 \pi \frac{\epsilon_0}{R^2} \, \Theta^N = 0 \, ,
\end{equation} 
under the following boundary conditions: 
\begin{equation}
\frac{d\omega_0(0)}{dr}=0 \, , \ \ \ \ 
\omega_0(R)+\frac{R}{3}\,\frac{d\omega_0(R)}{dr}=0  \, . 
\end{equation} 

In the Newtonian limit, $\frac{2M}{R}\rightarrow0$, eigensolutions of 
equation (\ref{b-eq}) should become those of $r$-mode oscillations in 
Newtonian stars. Therefore, eigenvalue $\bar{\sigma}$ and eigenfunction 
can be expanded in 
power series of $\frac{2M}{R}$ as 
\begin{equation}
\bar{\sigma}=\frac{2 m \Omega}{l(l+1)}\,\left[1+\kappa_1\,(\frac{2 M}{R}) +
O\left(\left(\frac{2 M}{R}\right)^2\right)\right] \, ,
\label{sig-exp}
\end{equation}
\begin{equation}
h_{0,l}=h^{(0)}_{0,l}\,\left(\frac{2 M}{R}\right) + 
O\left(\left(\frac{2 M}{R}\right)^2\right) \, ,
\label{ef-exp}
\end{equation}
and
\begin{equation}
U_l = U^{(0)}_l + O\left(\left(\frac{2 M}{R}\right)\right) \, .
\label{uu-exp}
\end{equation}
Substituting equations (\ref{pn1})-(\ref{pn5}) and (\ref{sig-exp}) into 
an equation (\ref{b-eq}), we obtain basic equations for $r$-mode oscillations 
in the post-Newtonian order as follows: 
\begin{equation}
(\kappa_1+\omega_0)\left[\frac{d^2 h^{(0)}_{0,l}(r)}{dr^2}-
\frac{l(l+1)}{r^2}\,h^{(0)}_{0,l}(r)\right] +
16\pi\,\frac{\epsilon_0}{R^2}\,\Theta^N\,h^{(0)}_{0,l}(r) = 0 \, ,
\end{equation}
and
\begin{equation}
U^{(0)}_l=-\,\frac{h^{(0)}_{0,l}}{\kappa_1+\omega_0} \, .
\end{equation}
These equations have the same properties as those of relativistic equations 
(\ref{b-eq}). Boundary conditions required for physically acceptable 
solutions are given as follows:
\begin{equation}
r\,\frac{d h^{(0)}_{0,l}}{dr}-(l+1)\,h^{(0)}_{0,l}=0 \ \ \ {\rm as} \ \  
r \rightarrow 0 \, ,
\end{equation}
and
\begin{equation}
r\,\frac{d h^{(0)}_{0,l}}{dr}+l h^{(0)}_{0,l}=0 \ \ \ {\rm at} \ \ r = R \, . 
\end{equation}
From inequalities (\ref{dis-sp}) and (\ref{con-sp}), frequency range for 
regular and singular eigenvalue problem in the post-Newtonian order 
equation are given as  
\begin{equation}
-\omega_0 (R) < \kappa_1 \le 0  \, ,
\label{pn-dis}
\end{equation}
and  
\begin{equation}
-\omega_0 (0) \le \kappa_1 \le -\omega_0 (R) \, ,
\label{pn-con}
\end{equation}
respectively.

Because our basic equation for both relativistic and 
post-Newtonian $r$-modes become second order linear 
ordinary differential equations with two boundary conditions, 
those are numerically solved as an eigenvalue problem, by 
using a Henyey type relaxation method (e.g., Unno et al. 1989).


\section{Numerical results} 

\subsection{Relativistic study}

As mentioned in the last section we should distinguish two cases in 
treating equation (\ref{b-eq}). One is a regular eigenvalue problem.
The other is a singular eigenvalue problem.
In this study we consider only $r$-mode solutions of a regular eigenvalue 
problem, because discrete eigenfrequency may be obtained. Furthermore such 
discrete modes are considered to correspond to Newtonian $r$-modes, 
because such modes have the well-defined regular eigenfunctions.   
Therefore we restrict our attention to 
eigenfrequency in the range (\ref{dis-sp}). For simplicity, henceforth, we 
will naively call a $r$-mode solution of a regular eigenvalue problem a 
discrete $r$-mode unless we give a notice. As mentioned in 
Beyer \& Kokkotas (1999) we cannot exclude the possibility that isolated 
eigenvalues may exist within a frequency range for a singular eigenvalue 
problem. Thus note that such an isolated eigenvalue does not belong to 
``discrete $r$-mode'' according to the terminology of this paper.

We compute frequencies of a discrete $r$-mode for several polytropic stellar 
models.  In present study, only the fundamental $r$-modes, whose 
eigenfunction $U_l$ has no node in radial direction except at 
stellar center, are obtained. This is consistent with that in Newtonian case, 
where overtone $r$-modes  appear as high order effects in angular velocity 
$\Omega$ (see, e.g. Provost, Berthomieu \& Rocca 1981, Saio 1982). 
In Figure 1, scaled eigenfrequencies 
$\kappa \equiv \bar{\sigma}/\Omega$ of 
discrete $r$-modes are given as functions of $M/R$ for modes with $m=l$. Note 
that for modes associated with $l$, we can directly obtain an eigenfrequency 
of a mode 
with any value of $m$ from that with $m=l$. Therefore only results for modes 
associated with $m=l$ will be shown throughout this paper. 
Eigenfrequencies for stars with four different polytropic indices, $N=0$, 
$0.5$, $0.75$, and $1$, are shown in panels in Figure 1, respectively. 
In each panel, frequency curves for 
the modes with $l=2$, $3$, and $4$ are depicted along a relativistic 
factor $M/R$. 
From the figure it is found that frequencies $\kappa$ monotonically decrease 
as a relativistic 
factor $M/R$ increases. This behavior is caused by dragging of 
inertial frame due to the stellar rotation. 
We find that some frequency curves in Figure 1 are terminated at 
some value of $M/R$ beyond which equilibrium states can still exist. 
Here, the maximum values of $M/R$ for polytropic equilibrium stars having 
$N=0$, $0.5$, $0.75$, and $1.0$ are given by $4/9$, $0.385$, $0.349$, and 
$0.312$, respectively. 
It is also found that length of those curves tend to be short 
as either a polytrope index $N$ or an angular quantum number $l$ increases. 
As a matter of fact, we are not able to obtain eigenfrequencies beyond them
by using our numerical method.

In order to trace the cause that missing frequencies in Figure 1 cannot be  
obtained, let us see behavior of values of 
$\frac{2 m \bar{\omega}(R)}{l(l+1)\Omega}$ as a function of $M/R$. 
The value of 
$\frac{2 m \bar{\omega}(R)}{l(l+1)\Omega}$ gives a lower bound of 
the eigenfrequency $\kappa$ allowed for the regular eigenvalue problem of a 
$r$-mode in the considered star. Therefore an eigenfrequency $\kappa$ for 
the regular eigenvalue problem must be 
larger than $\frac{2 m \bar{\omega}(R)}{l(l+1)\Omega}$. 
In Figure 2, values of 
$\frac{2 m \bar{\omega}(R)}{l(l+1)\Omega}$ and $\kappa$ for $l=m=2$ 
$r$-mode of $N=0.75$ polytropic stars are given as function of $M/R$. 
In this figure it is found that the curve of $\kappa$ approaches that of 
$\frac{2 m \bar{\omega}(R)}{l(l+1)\Omega}$ twice, that is, 
near the Newtonian limit, and near the terminal 
point of the frequency curve. 
The former is clear because the frequency range (\ref{dis-sp}) shrinks
to the point of the Newtonian r-mode frequency.
The latter is considered to be the cause of our problem. 
From the figure it is found that frequency $\kappa$ of $r$-mode 
solutions for the regular eigenvalue problem becomes the same value as 
that of $\frac{2 m \bar{\omega}(R)}{l(l+1)\Omega}$ at some critical 
value of $M/R$, say $(M/R)_{crit}$, within the numerical accuracy.  
In other words,  
$\kappa$ nearly becomes the frequency for the singular eigenvalue problem 
as the value of $M/R$ becomes $(M/R)_{crit}$. Thus we can not obtain those 
missing modes. Behaviors of $\frac{2 m \bar{\omega}(R)}{l(l+1)\Omega}$ and 
$\kappa$ as functions of $M/R$ for 
other cases except for the $N=0$ case are similar to those in Figure 2. 
As for the $N=0$ case, the curve of $\kappa$ does not closely approach that of 
$\frac{2 m \bar{\omega}(R)}{l(l+1)\Omega}$ even when the value of $M/R$ 
is nearly $4/9$.  
From those facts and continuity of physical quantities we can expect that 
disappearance of $r$-mode solutions with frequency in a range (\ref{dis-sp})
is not due to numerical artifact. 
Thus it is reasonable to claim that $r$-mode solutions for regular 
eigenvalue problem disappear in stars having a values of $M/R$ larger 
than $(M/R)_{crit}$.

Critical values of $M/R$ are shown as functions of polytropic index $N$ 
in Figure 3. Curves of critical values 
$(M/R)_{crit}$ for modes with $l=2$, $3$, and $4$ are given in the figure. 
Note that values of $(M/R)_{crit}$ shown here are numerically approximated 
ones. These only give lower limit of accurate values of $(M/R)_{crit}$. 
In Figure 3, stars whose parameters are in regions under curves of 
$(M/R)_{crit}$ have discrete $r$-mode spectrum. 
Results of our numerical analysis also suggest that 
for the $N=0$ case the critical value of $M/R$ is nearly equal to $4/9$. 
Indeed, discrete $r$-mode solutions with frequency in a range (\ref{dis-sp}) 
may exist in all uniform density stars independently of the value of $M/R$. 
This is consistent with the result by Lockitch et al. (2001). 
The figure also suggests that even in nearly Newtonian stars, whose value 
of relativistic factor $M/R$ is sufficiently small, $r$-mode solutions of 
a regular eigenvalue problem disappear if a value of $N$ is over some 
critical value, say $N_{crit}$. For such nearly Newtonian stars, 
however, our numerical method tends to lose an accuracy. In the next 
subsection, 
to avoid this numerical difficulty, we will study $r$-mode oscillations of 
nearly Newtonian stars by using post-Newtonian formalism.

\subsection{Post-Newtonian study}

In post-Newtonian analysis of $r$-modes, the post-Newtonian correction 
$\kappa_1$ for the eigenfrequency is employed as the eigenvalue instead of 
a scaled eigenfrequency $\kappa$. The value of $\kappa_1$ does not depend 
on a relativistic factor $M/R$ by definition (\ref{sig-exp}), 
but both on polytropic index $N$ and on angular eigenvalue $l$. 
In similar way to the case of the relativistic study we consider only the 
case that our basic equations become the regular eigenvalue problem. Thus we 
restrict eigensolutions to those with frequency in a range (\ref{pn-dis}). 
In Figure 4 the post-Newtonian correction $\kappa_1$ are plotted against 
polytropic index $N$ for three different eigenmodes associated with $l=2$, 
$3$ and $4$. Our numerical results of $\kappa_1$ for a $N=0$ polytropic 
model are in good agreement with those by Lockitch et al. (2001). 
We also cannot obtain $r$-mode solutions for the regular 
eigenvalue problem in stars beyond some limits of polytropic 
index $N$, as shown in the figure. Thus as in the relativistic case,
examining behaviors of values of 
$-\omega_0(R)$ appearing in the frequency inequality (\ref{pn-dis}) as a 
function of $N$, we confirm that 
disappearance of $r$-mode solutions with frequency in a range (\ref{pn-dis}) 
is not due to numerical artifact. The values of $-\omega_0(R)$ are shown as 
function of $N$ in Figure 4. It is found that all frequency curves for 
$\kappa_1$ in Figure 4 nearly intersect with the curve for $-\omega_0(R)$ 
at some values of $N$, namely $N = N_{crit}$. In the same way as the 
relativistic 
study, thus, we can conclude that the absence of the post-Newtonian 
$r$-mode solutions for the regular eigenvalue problem in stars having 
$N > N_{crit}$ is not due to numerical artifact.
From this  we expect 
that $r$-mode solutions with a frequency in a range (\ref{pn-dis}) 
disappear if a star has polytrope index $N$ larger than a 
critical value $N_{crit}$. Values of $N_{crit}$ obtained from 
numerical analysis are shown in Figure 4 near the terminal points of 
$\kappa_1$ curves. Our numerical results of $N_{crit}$ in the 
post-Newtonian order are consistent with those in relativistic 
analysis. (compare with Figure 3)
From the figure, for instance, we can expect that the
$r$-mode solution with $l=2$ for regular eigenvalue problem disappears in 
polytope having $N>1.18$ even in the post-Newtonian order.


\section{Discussion and Conclusion}

In this paper, we have investigated the properties of $r$-modes characterized
by regular eigenvalue problem in a slowly rotating relativistic polytrope.
Our numerical results suggest that discrete $r$-mode solutions with
frequency in a range (\ref{dis-sp}) disappear for some polytropic models.
Although we do not have mathematically rigorous proof for this suggestion,
from numerical results and continuity of physical quantities we can
make conjectures on the relativistic $r$-mode as follows: the $r$-mode
associated with $l=m=2$, which is considered to be the most important for
gravitational radiation driven instability, do not
exist as a solution for regular eigenvalue problem for polytropes having
$N > 1.18$ even in the post-Newtonian order. As for a $N=1$ polytropic
model, which is employed as a typical neutron star model, discrete $r$-mode
solutions with frequency in a range (\ref{dis-sp}) do not exist for stars
whose relativistic factor $M/R$ is larger than about $0.1$.
If those conjectures are true, most of typical neutron stars do not have
discrete $r$-mode solution with frequency in a range (\ref{dis-sp}),
because a typical
relativistic factor for a neutron star is considered to be about
$0.1$--$0.2$.  This means that in actual neutron stars $r$-mode
characterized by regular eigenvalue problem cannot play any important role,
if relativistic $r$-mode oscillations are governed by equation (\ref{b-eq}).
If this is the case, we must study the pure $r$-mode with frequency
in a range (\ref{con-sp}) to have clear understanding of relativistic
$r$-mode in neutron stars. According to previous studies,
continuous part of spectrum and
singular eigenfunctions appear for such a frequency range, although
we cannot exclude the possibility that isolated eigenvalues may exist
within the frequency range.

In this study, we employ the formalism developed by Lockitch et al. (2001) 
to investigate $r$-modes in relativistic polytropes. Although their 
formalism is correct, the system of equation (\ref{b-eq}) as regular
eigenvalue problem leads to some unnatural
numerical results. For instance, $l=2$ $r$-mode solutions of regular
eigenvalue problem in $N=1$ polytropes having $M/R > 0.1$ disappear due to
general relativistic effect, although $r$-mode solutions for regular
eigenvalue problem in Newtonian stars exist for all slowly rotating
polytropic stars with constant angular velocity. 
Lockitch et al. (2001) claimed that $r$-mode solutions with frequency in 
a range (\ref{dis-sp}) are counterpart of Newtonian $r$-modes, because 
they considered that the true relativistic analogue to a Newtonian $r$-mode
should be a distinct mode with a well-defined eigenfunction. 
In fact, such a $r$-mode solution may
exist in all $N=0$ polytrope, as shown numerically in Lockitch et al. (2001)
and this study. However, for some $N\ne 0$ polytropes we numerically found
that $r$-mode solutions with frequency in a range (\ref{dis-sp}) do not appear
when $M/R > (M/R)_{crit}$ or $N > N_{crit}$. We think that this is actual 
behavior of the mode due to the
influence of the effective differential rotation caused by general
relativity. The reason of this is considered as follows:
In modal analysis in Newtonian rotating stars,
sufficient differential rotation of an equilibrium star can give a
corotation point, where corotation frequency $\bar{\sigma}$ vanishes. Because
of existence of this corotation point the system may become singular 
eigenvalue problem. (see, e.g. Karino, Yoshida \& Eriguchi 2000) On the other 
hand, our results
show that disappearance of discrete r-mode occurs
in large values of the relativistic factor or
the polytropic index of the equilibrium star. In such equilibrium stars, as
a matter of fact,
differential rotation effect in relativistic rotating stars becomes strong.

However we cannot exclude the possibility that pure $r$-modes in a
relativistic rotating star with the non-zero rotation frequency are not
simply governed by equation (\ref{b-eq}), because strictly speaking the 
equation (\ref{b-eq}) only determines zero frequency modes in a spherical 
non-rotating star. In other worlds, in equation (\ref{b-eq}) the 
gravitational radiation reaction 
is not taken into account at all, although the gravitational radiation 
must be emitted from $r$-mode oscillations in rotating stars with non-zero 
rotation velocity.  
As strongly suggested by Lockitch et al. (2001) (see, also Beyer \& 
Kokkotas 1999), when
the eigenfrequency becomes complex number, which expresses the damping of
the oscillation due to the energy dissipation such as the gravitational
radiation emission, the existence of the continuous part of spectrum
in the eigenfrequency of equation (\ref{b-eq}) might be avoided.
If this is true, there is a possibility for a distinct mode 
solution with a well-defined eigenfunction from singular solutions with 
frequency in a range (\ref{con-sp}) to be picked up by the information due to 
the gravitational radiation reaction effect. Such distinct modes
may make a continuous sequence that starts with $r$-mode solutions
with frequency in a range (\ref{dis-sp}). We think this one of the 
most plausible scenarios, although this is still speculation. In order to 
have clear understanding of the $r$-mode in relativistic stars, the 
further progress of investigations of the rotational mode is needed.


\acknowledgements

The author would like to thank Prof. B. F. Schutz for his hospitality at 
the Albert Einstein Institute, where a part of this work was done. 
He is grateful to Prof. T. Futamase and Dr. S'i Yoshida for useful comments 
and careful reading of the manuscript. He also would like to thank 
Drs. H. R. Beyer and U. Lee for valuable comments.


%


\newpage

\begin{figure}
\epsscale{.8}
\plotone{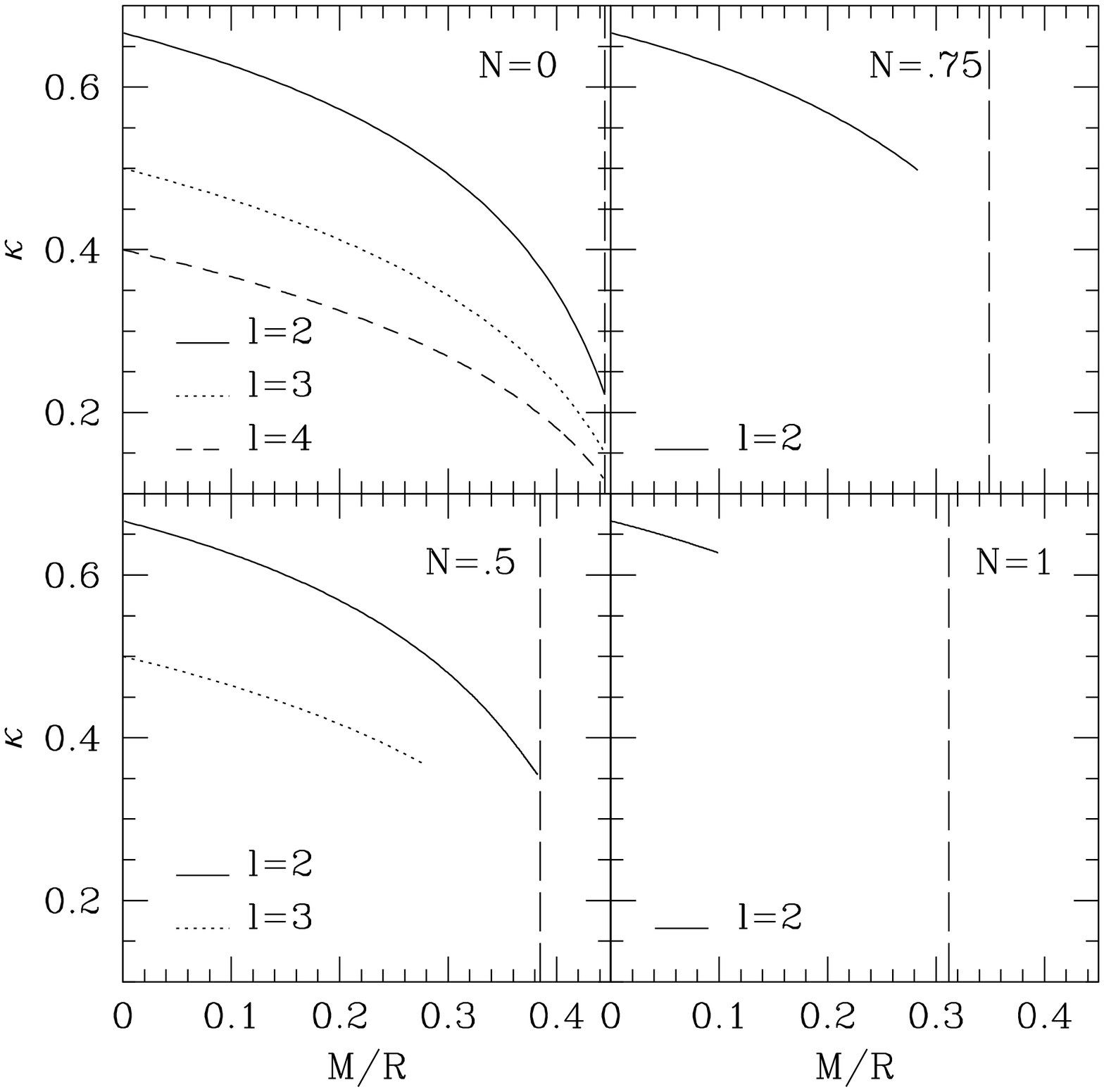}
\caption{Scaled frequencies $\kappa=\bar{\sigma}/\Omega$ of the 
$r$-modes in polytropic models with $N=0$, $0.5$, $0.75$, and $1$ are 
plotted as functions of $M/R$ in panels, respectively. In each panel, 
the frequencies for modes with $l=2$, $3$, and $4$ are given. The labels 
indicating their polytropic index $N$ are attached in corresponding panels.  
The eigenvalue of the $r$-modes that is described above but not 
given in the figure, for instance $l=3$ mode in $N=1$ polytrope, are not  
obtained by using our numerical method. Vertical dotted lines show the 
maximum values of $M/R$ for equilibrium states: $M/R=0.444$ for $N=0$, 
$M/R=0.385$ for $N=0.5$, $M/R=0.349$ for $N=0.75$, and 
$M/R=0.312$ for $N=1.0$.}
\end{figure}

\begin{figure}
\epsscale{.8}
\plotone{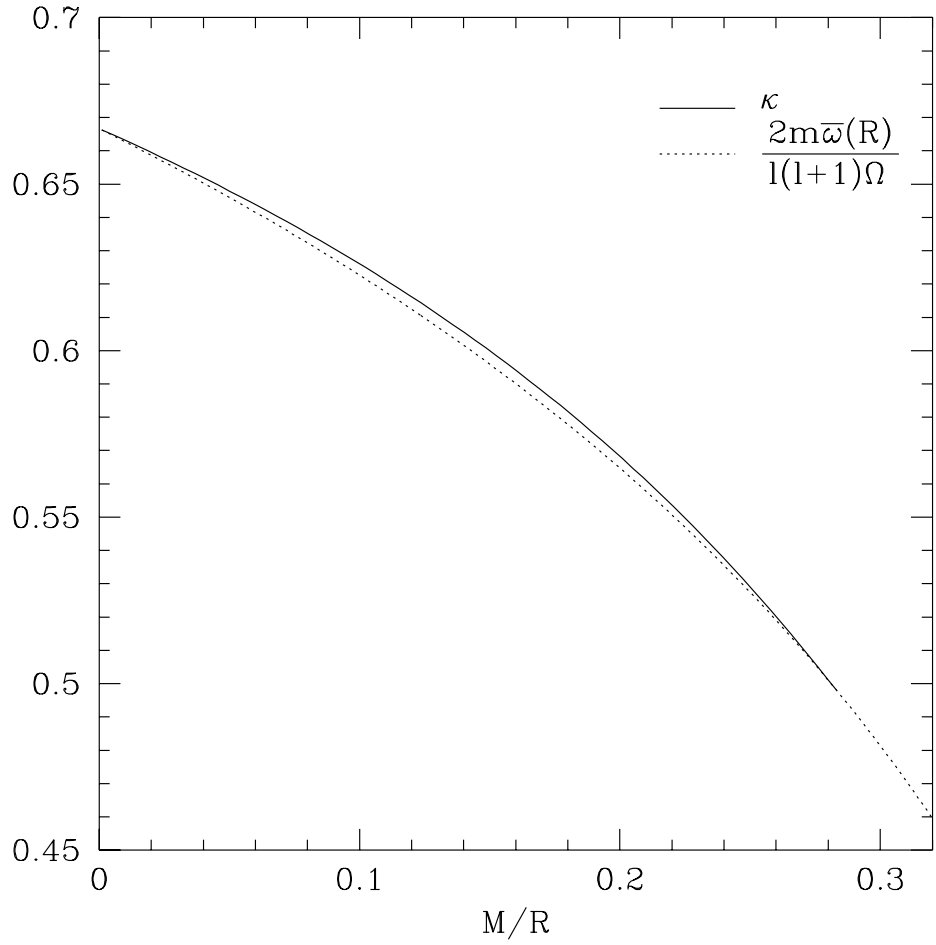}
\caption{The values of 
${2m\bar{\omega}(R) \over l(l+1)\Omega}$ and $\kappa$ for $l=2$ 
$r$-modes of $N=0.75$ polytropes are plotted as functions of $M/R$.}
\end{figure}

\begin{figure}
\epsscale{.8}
\plotone{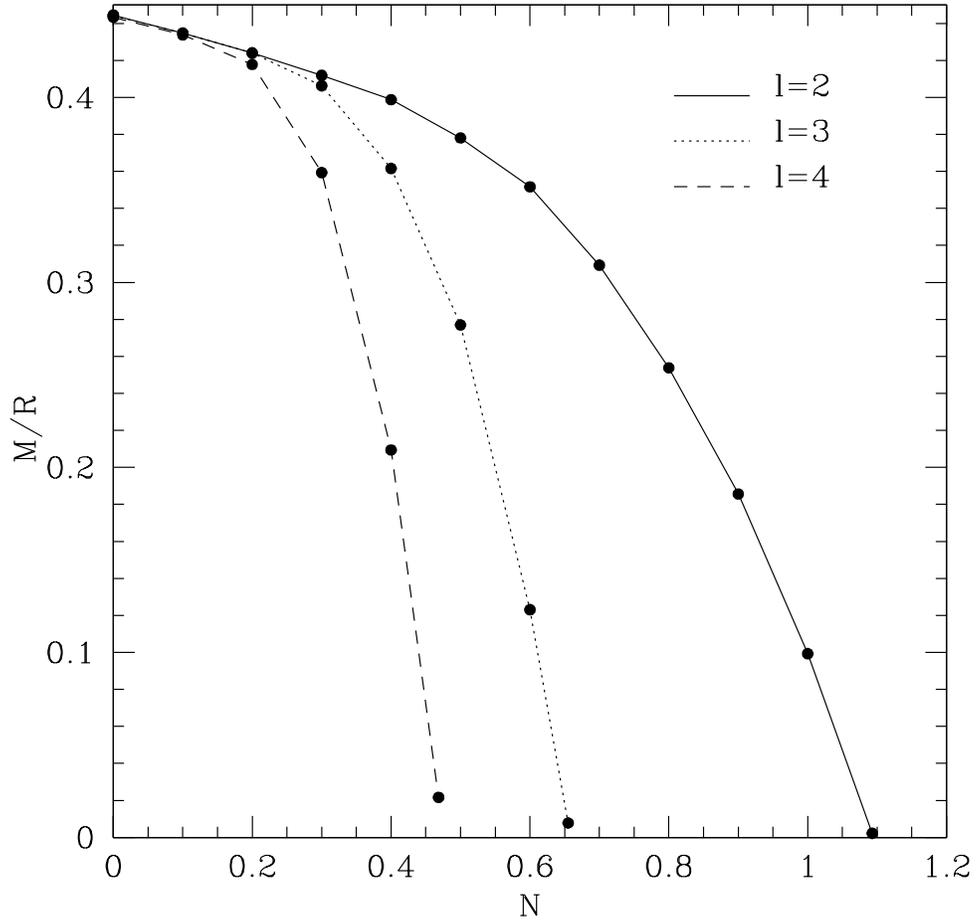}
\caption{The critical values of relativistic factor $M/R$ are 
plotted against the polytropic index $N$ for discrete $r$-modes with $l=2$, 
$3$, and $4$. The values of $(M/R)_{crit}$ obtained from actual numerical 
calculations are indicated by the filled circles.}  
\end{figure}

\begin{figure}
\epsscale{.8}
\plotone{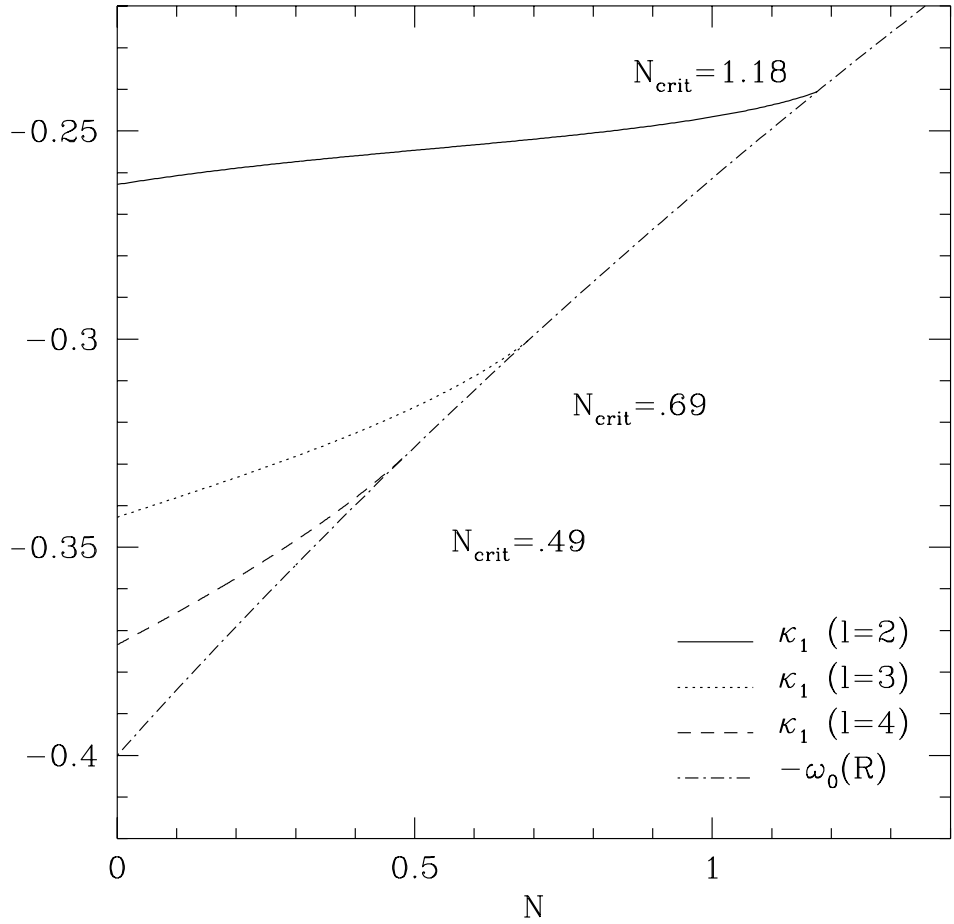}
\caption{The post-Newtonian correction $\kappa_1$ for discrete 
$r$-mode frequencies with $l=2$, $3$, and $4$ are shown as functions of the 
polytropic index $N$. The labels indicating the critical values of $N$ are 
attached near the terminal points for each frequency curves. The value of 
$-\omega_0(R)$ is also plotted as a function of $N$.}
\end{figure}

\end{document}